\documentclass[conference]{IEEEtran}
\IEEEoverridecommandlockouts
\usepackage{cite}
\usepackage{amsmath,amssymb,amsfonts}
\usepackage{algorithmic}
\usepackage{graphicx}
\usepackage{textcomp}
\usepackage{xcolor}
\usepackage{booktabs} 
\def\BibTeX{{\rm B\kern-.05em{\sc i\kern-.025em b}\kern-.08em
    T\kern-.1667em\lower.7ex\hbox{E}\kern-.125emX}}
\begin{document}

\title{Quantum Kernel-Based Long Short-term Memory\\

\thanks{\IEEEauthorrefmark{1} Corresponding Author: kuan-cheng.chen17@imperial.ac.uk}
}

\author{
\IEEEauthorblockN{
    Yu-Chao Hsu\IEEEauthorrefmark{2}\IEEEauthorrefmark{3},
    Tai-Yu Li\IEEEauthorrefmark{4}, 
    Kuan-Cheng Chen\IEEEauthorrefmark{5}\IEEEauthorrefmark{6}\IEEEauthorrefmark{1}
}
\IEEEauthorblockA{\IEEEauthorrefmark{2} National Center for High-Performance Computing, NARlabs, Hsinchu, Taiwan}
\IEEEauthorblockA{\IEEEauthorrefmark{3} Cross College Elite Program, National Cheng Kung University, Tainan, Taiwan}
\IEEEauthorblockA{\IEEEauthorrefmark{4}National Synchrotron Radiation Research Center, Hsinchu, Taiwan}
\IEEEauthorblockA{\IEEEauthorrefmark{5}Department of Electrical and Electronic Engineering, Imperial College London, London, UK}
\IEEEauthorblockA{\IEEEauthorrefmark{6}Centre for Quantum Engineering, Science and Technology (QuEST), Imperial College London, London, UK}

}

\maketitle

\begin{abstract}
The integration of quantum computing into classical machine learning architectures has emerged as a promising approach to enhance model efficiency and computational capacity. In this work, we introduce the Quantum Kernel-Based Long Short-Term Memory (QK-LSTM) network, which utilizes quantum kernel functions within the classical LSTM framework to capture complex, non-linear patterns in sequential data. By embedding input data into a high-dimensional quantum feature space, the QK-LSTM model reduces the reliance on large parameter sets, achieving effective compression while maintaining accuracy in sequence modeling tasks. This quantum-enhanced architecture demonstrates efficient convergence, robust loss minimization, and model compactness, making it suitable for deployment in edge computing environments and resource-limited quantum devices (especially in the NISQ era). Benchmark comparisons reveal that QK-LSTM achieves performance on par with classical LSTM models, yet with fewer parameters, underscoring its potential to advance quantum machine learning applications in natural language processing and other domains requiring efficient temporal data processing.
\end{abstract}


\section{Introduction}
Sequence modeling tasks, including natural language processing (NLP), time series forecasting, and signal classification, are pivotal in numerous domains of computer science and engineering. Recurrent Neural Networks (RNNs)\cite{RNN} and Long Short-Term Memory (LSTM) \cite{LSTM} networks have been instrumental in addressing these tasks due to their capability to capture temporal dependencies within sequential data. However, as the complexity and dimensionality of data continue to escalate, classical RNNs and LSTMs often demand substantial computational resources and extensive parameterization to effectively model intricate patterns and long-range dependencies\cite{yu2019review}.

Quantum computing has emerged as a promising paradigm that leverages quantum mechanical principles such as superposition and entanglement to enhance machine learning models, offering significant speed advantages over traditional computation\cite{huang2021power}. Specifically, quantum machine learning (QML) aims to exploit the computational advantages of quantum systems to process information in high-dimensional Hilbert spaces more efficiently than classical counterparts\cite{peters2021machine,biamonte2017quantum,yu2024shedding}. This capability positions quantum computing advantageously for large-scale and high-dimensional applications, including high-energy physics\cite{guan2021quantum,di2024quantum,wu2021application}, medical science\cite{emani2021quantum,chen2024compressedmediq,li2022classification}, signal processing\cite{chen2020variational,yang2021decentralizing,lin2024quantum}, climate change\cite{ho2024quantum,nammouchi2023quantum}, cosmology\cite{chen2024quantumICIC}, NLP\cite{di2022dawn} and finance\cite{orus2019quantum,cao2023linear}. In the realm of time series prediction, prior efforts to integrate quantum computing into sequence modeling have led to the development of Quantum-Enhanced Long Short-Term Memory (QLSTM)\cite{chen2022quantum} and Quantum-Trained LSTM\cite{lin2024quantum,liu2024qtrl} architectures based on Variational Quantum Circuit (VQC)\cite{chen2020variational}. Although VQC-based QLSTMs incorporate quantum circuits into neural network structures, they often involve complex circuit designs and require substantial quantum resources, posing significant challenges for implementation on current quantum hardware\cite{preskill2018quantum}. 

In contrast, quantum kernel methods offer an alternative approach by embedding classical data into quantum feature spaces using quantum circuits\cite{blank2020quantum}, enabling efficient computation of inner products (kernels) in these high-dimensional spaces\cite{rebentrost2014quantum,li2015experimental}. Quantum kernels can capture complex data structures with potentially fewer trainable parameters and reduced computational overhead compared to both classical models and VQC-based quantum models\cite{maheshwari2021variational}. This approach leverages the ability of quantum systems to represent and manipulate high-dimensional data efficiently, providing a pathway to enhance model expressiveness without proportionally increasing computational demands\cite{gentinetta2024complexity}.

This paper introduces the QK-LSTM network, which integrates quantum kernel computations within the LSTM architecture to enhance the modeling of complex sequential patterns. By replacing classical linear transformations in the LSTM cells with quantum kernel evaluations, the QK-LSTM leverages quantum feature spaces to encode intricate dependencies more effectively. This approach harnesses quantum gates and circuits to perform transformations that would be computationally intensive in classical settings, thereby enhancing the efficiency of the network. Moreover, this integration simplifies the quantum circuit requirements compared to  VQC-based QLSTMs, making the QK-LSTM more feasible for implementation on near-term quantum devices and suitable for deployment in quantum edge computing\cite{ma2022hybrid} and resource-constrained environments. Additionally, the quantum kernel can serve as an effective ansatz for distributed quantum computing, suggesting that this method can be extended towards quantum HPC and distributed quantum computing architectures\cite{chen2024consensus,chen2024cutn,burt2024generalised}.

\section{Method}
\subsection{Long Short-Term Memory}

LSTM networks \cite{LSTM} are a specialized form of RNNs \cite{RNN}, particularly adept at capturing extended sequential dependencies in data. When applied to Part-of-Speech (POS) tagging tasks \cite{pos2000,pos2009,pos2015}, the LSTM model processes each word in a sentence sequentially, leveraging its memory cells to retain contextual information. This approach allows it to assign the correct POS tag to each word by considering both past and future context within the sequence.

\begin{figure}[!b]
    \centering
    \includegraphics[width=0.4\textwidth]{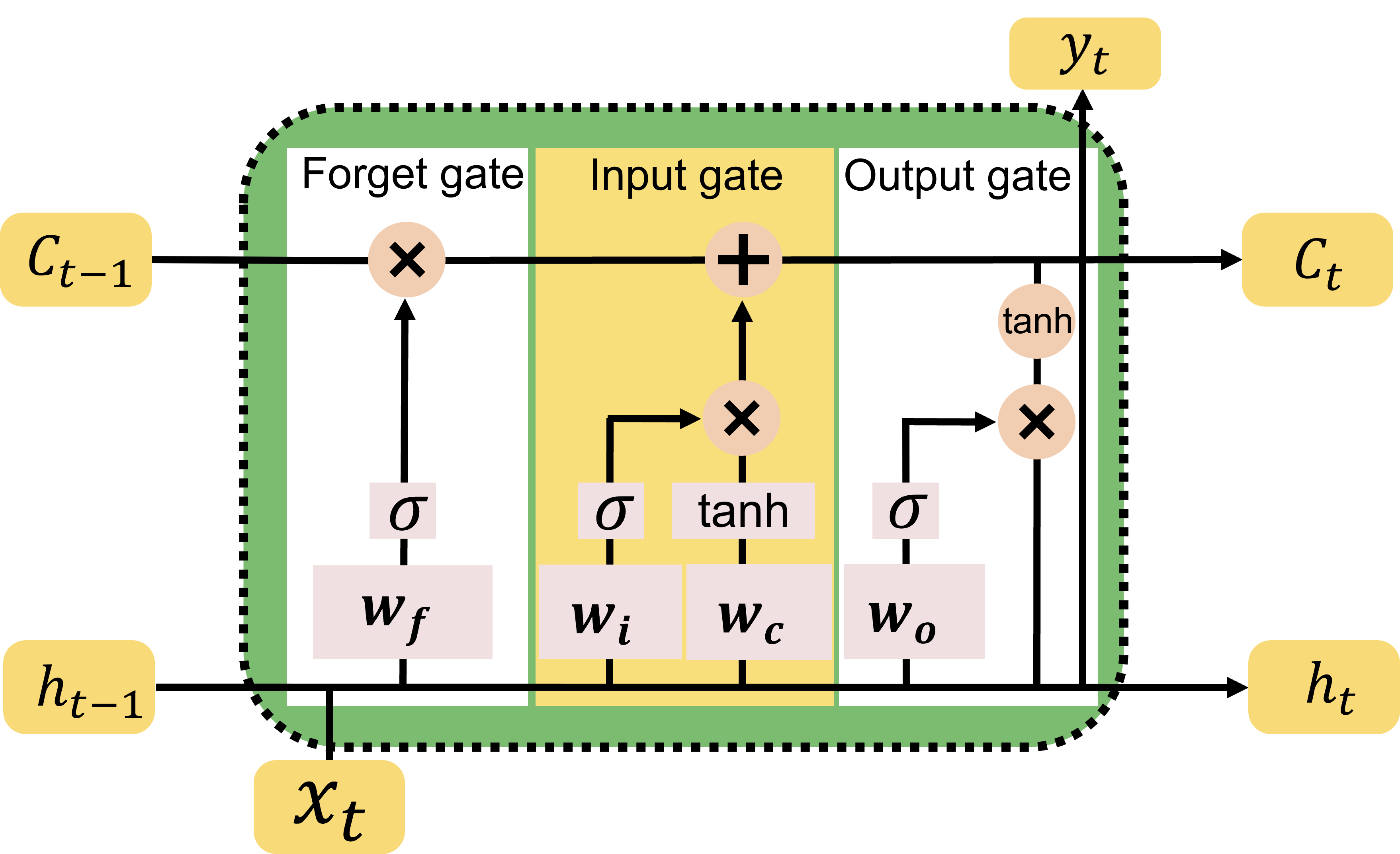}
    \caption{Schematic representation of a standard classical LSTM cell.}
    \label{LSTM}
\end{figure}

Unlike traditional methods such as Hidden Markov Models (HMMs) \cite{HMM} and Conditional Random Fields (CRFs) \cite{CRF}, LSTM networks can capture long-range dependencies due to their unique gating mechanisms. This capability enhances their understanding of syntactic patterns in complex sentences, establishing LSTM as a powerful tool for NLP tasks, including POS tagging. A schematic representation of a standard classical LSTM cell is illustrated in Fig.~\ref{LSTM}.

\subsection{Quantum Kernel-Based LSTM}

\begin{figure*}[!t]
    \centering
    \includegraphics[width=0.75\textwidth]{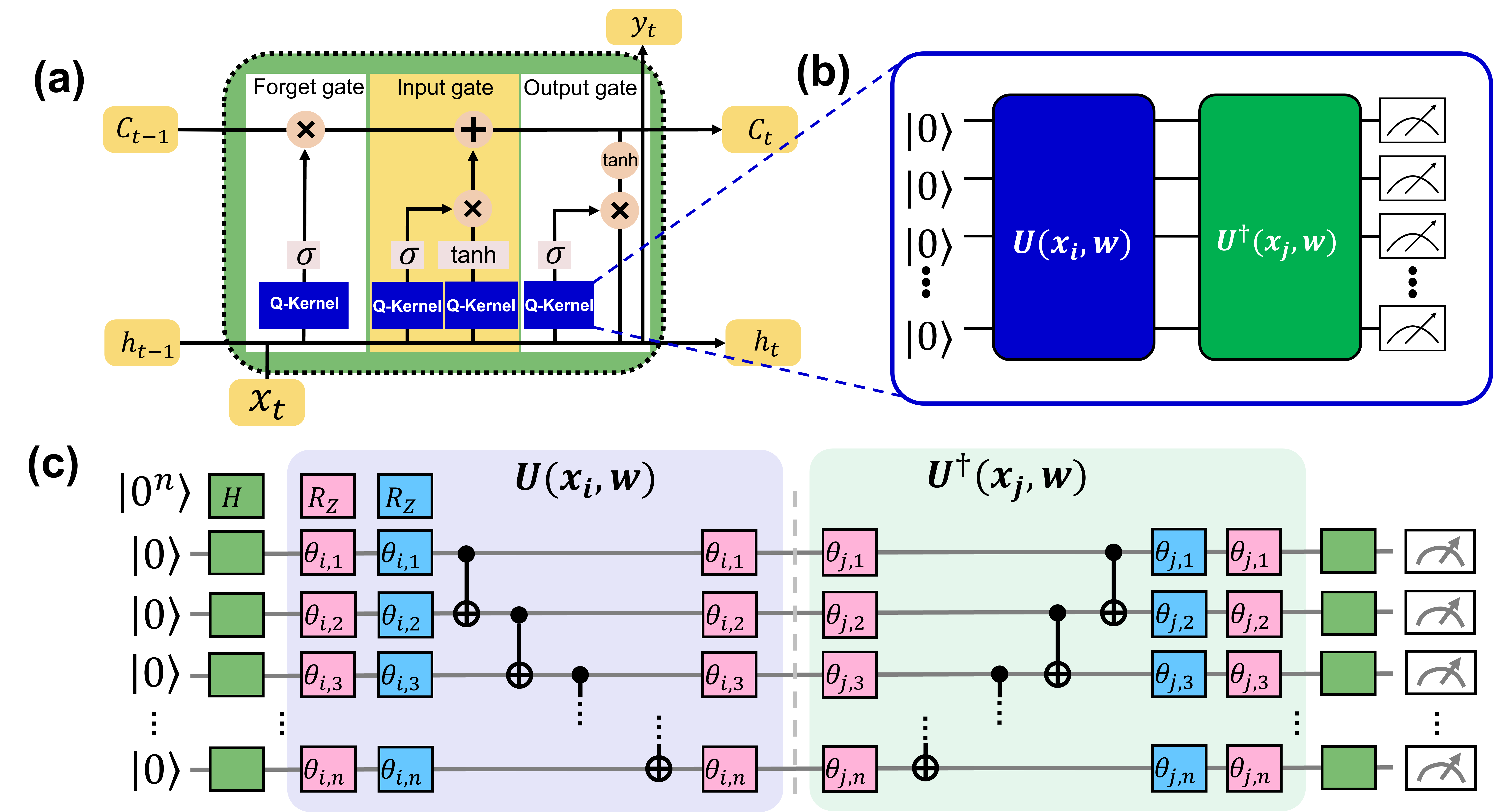}
    \caption{Overview of the QK-LSTM Architecture. (a) The QK-LSTM cell integrates quantum kernel transformations within the conventional LSTM framework, where each gate (forget, input, and output) utilizes quantum kernels to enhance sequential data processing and retain temporal dependencies. (b) The unitary gate representation of the quantum kernel, denoted as \(U(x_i, w)\), maps classical input data \(x_t\) into a quantum feature space, with the conjugate transpose \(U^{\dagger}(x_j, w)\) facilitating quantum state overlap calculations. (c) The full quantum circuit of the QSVM, which applies quantum kernel-based transformations to encode data, aiding in quantum-enhanced machine-learning tasks within the QK-LSTM model.}
    \label{QK-LSTM}
\end{figure*}

In this part, we introduce the Quantum Kernel-Based Long Short-Term Memory (QK-LSTM) architecture, which integrates quantum kernel computations into the classical LSTM framework to enhance its ability to capture complex, non-linear patterns in sequential data.

As illustrated in Fig.~\ref{QK-LSTM}, the fundamental unit of the proposed QK-LSTM architecture is the QK-LSTM cell. Each QK-LSTM cell modifies the standard LSTM cell by replacing the linear transformations with quantum kernel evaluations, effectively embedding the input data into a high-dimensional quantum feature space.

\subsubsection{Classical LSTM}

The standard LSTM cell comprises three gates—the forget gate \( f_t \), the input gate \( i_t \), and the output gate \( o_t \)—and the cell state \( C_t \). The classical LSTM equations are:

\begin{subequations}
\begin{align}
f_t &= \sigma\left( W_f [h_{t-1}, x_t] + b_f \right), \label{eq:classical_forget_gate} \\
i_t &= \sigma\left( W_i [h_{t-1}, x_t] + b_i \right), \label{eq:classical_input_gate} \\
\tilde{C}_t &= \tanh\left( W_C [h_{t-1}, x_t] + b_C \right), \label{eq:classical_candidate_cell_state} \\
C_t &= f_t \odot C_{t-1} + i_t \odot \tilde{C}_t, \label{eq:classical_cell_state_update} \\
o_t &= \sigma\left( W_o [h_{t-1}, x_t] + b_o \right), \label{eq:classical_output_gate} \\
h_t &= o_t \odot \tanh\left( C_t \right), \label{eq:classical_hidden_state}
\end{align}
\end{subequations}

where:
- \( x_t \) is the input vector at time \( t \),
- \( h_{t-1} \) is the hidden state from the previous time step,
- \( W \) and \( b \) are weight matrices and biases,
- \( \sigma \) denotes the sigmoid activation function,
- \( \tanh \) denotes the hyperbolic tangent activation function,
- \( \odot \) denotes element-wise multiplication.

\subsubsection{Quantum Kernel Integration into LSTM}

In the QK-LSTM architecture, we replace the linear transformations \( W [h_{t-1}, x_t] + b \) in the gate computations with quantum kernel evaluations. The idea is to leverage the expressive power of quantum feature spaces to model complex, non-linear relationships in the data.

Define the concatenated input vector:
\begin{equation}
v_t = [h_{t-1}, x_t].
\end{equation}

We introduce a set of reference vectors \( \{ v_j \}_{j=1}^{N} \), which are either a subset of training data or learned during training. The gate activations are computed using weighted sums of quantum kernel functions:

\begin{subequations}
\begin{align}
f_t &= \sigma\left( \sum_{j=1}^{N} \alpha_j^{(f)} k^{(f)}(v_t, v_j) + b_f \right), \label{eq:qk_forget_gate} \\
i_t &= \sigma\left( \sum_{j=1}^{N} \alpha_j^{(i)} k^{(i)}(v_t, v_j) + b_i \right), \label{eq:qk_input_gate} \\
\tilde{C}_t &= \tanh\left( \sum_{j=1}^{N} \alpha_j^{(C)} k^{(C)}(v_t, v_j) + b_C \right), \label{eq:qk_candidate_cell_state} \\
C_t &= f_t \odot C_{t-1} + i_t \odot \tilde{C}_t, \label{eq:qk_cell_state_update} \\
o_t &= \sigma\left( \sum_{j=1}^{N} \alpha_j^{(o)} k^{(o)}(v_t, v_j) + b_o \right), \label{eq:qk_output_gate} \\
h_t &= o_t \odot \tanh\left( C_t \right). \label{eq:qk_hidden_state}
\end{align}
\end{subequations}

Here:
- \( \alpha_j^{(f)} \), \( \alpha_j^{(i)} \), \( \alpha_j^{(C)} \), and \( \alpha_j^{(o)} \) are trainable weights associated with the quantum kernels for each gate,
- \( k^{(f)} \), \( k^{(i)} \), \( k^{(C)} \), and \( k^{(o)} \) are quantum kernel functions specific to each gate,
- \( b_f \), \( b_i \), \( b_C \), and \( b_o \) are biases.

\subsubsection{Quantum Kernel Function}

The quantum kernel function \( k(v_t, v_j) \) measures the similarity between two data points \( v_t \) and \( v_j \) in a quantum feature space induced by a quantum feature map \( \phi(v) \): \(k(v_t, v_j) = \left| \langle \phi(v_t) | \phi(v_j) \rangle \right|^2\).

The quantum feature map \( \phi(v) \) is implemented via a parameterized quantum circuit \( U(v) \) that encodes the classical data \( v \) into a quantum state \( | \phi(v) \rangle = U(v) | 0 \rangle^{\otimes n} \).

\paragraph{Quantum Circuit Design}

The quantum circuit \( U(v) \) consists of the following components:

1. \textbf{Initialization}: All qubits are initialized to the \( |0\rangle \) state.

2. \textbf{Hadamard Gates}: Apply Hadamard gates to create a superposition:
   \begin{equation}
   | \psi_0 \rangle = H^{\otimes n} | 0 \rangle^{\otimes n}.
   \end{equation}
   
3. \textbf{Data Encoding}: Encode classical data using parameterized rotation gates:
   \begin{equation}
   U_{\text{enc}}(v) = \prod_{k=1}^{n} R_y(\theta_{k}) R_z(\phi_{k}),
   \end{equation}
   where \( \theta_{k} \) and \( \phi_{k} \) are functions of the components of \( v \).
   
4. \textbf{Entanglement}: Introduce entanglement using CNOT gates:
   \begin{equation}
   U_{\text{ent}} = \prod_{k=1}^{n-1} \text{CNOT}(k, k+1).
   \end{equation}
   
5. \textbf{Final State}: The quantum state is:
   \begin{equation}
   | \phi(v) \rangle = U_{\text{ent}} U_{\text{enc}}(v) H^{\otimes n} | 0 \rangle^{\otimes n}.
   \end{equation}

\paragraph{Quantum Kernel Evaluation}

The quantum kernel between \( v_t \) and \( v_j \) is computed as:
\begin{equation}
k(v_t, v_j) = \left| \langle 0 |^{\otimes n} U^\dagger(v_j) U(v_t) | 0 \rangle^{\otimes n} \right|^2.
\end{equation}

This computation involves preparing the quantum states corresponding to \( v_t \) and \( v_j \), applying the inverse circuit \( U^\dagger(v_j) \) followed by \( U(v_t) \), and measuring the probability of the system being in the \( | 0 \rangle^{\otimes n} \) state.

\subsubsection{Training and Optimization}

The parameters of the QK-LSTM model include the weights \( \alpha_j \), biases \( b \), and any parameters within the quantum circuits used for the kernel computations.

\paragraph{Loss Function}

For a given task (e.g., classification or regression), we define a suitable loss function \( \mathcal{L} \). For example, for classification: \(L = 1/T \sum_{t=1}^{T} \mathcal{L}(y_t, \hat{y}_t)\), where \( y_t \) is the true label, \( \hat{y}_t \) is the predicted output, and \( T \) is the total number of time steps.

\paragraph{Gradient Computation}

The gradients of the loss with respect to the classical parameters \( \alpha_j \) and \( b \) are computed using standard backpropagation through time (BPTT). For the quantum circuit parameters, we employ the parameter-shift rule \cite{parameter-shift}, which allows efficient computation of gradients in quantum circuits.

\paragraph{Parameter-Shift Rule}

The gradient of the quantum kernel with respect to a circuit parameter \( \theta \) is given by:
\begin{equation}
\frac{\partial k(v_t, v_j)}{\partial \theta} = k_{\theta}^{+}(v_t, v_j) - k_{\theta}^{-}(v_t, v_j),
\end{equation}
where \( k_{\theta}^{\pm}(v_t, v_j) \) is the kernel evaluated with the parameter \( \theta \) shifted by \( \pm \frac{\pi}{2} \):
\begin{equation}
k_{\theta}^{\pm}(v_t, v_j) = \left| \langle 0 |^{\otimes n} U^\dagger(v_j) U_{\theta}^{\pm}(v_t) | 0 \rangle^{\otimes n} \right|^2.
\end{equation}

\paragraph{Optimization Algorithm}

An optimization algorithm such as stochastic gradient descent (SGD) or Adam is used to update the parameters:
\begin{align}
\alpha_j &\leftarrow \alpha_j - \eta \frac{\partial L}{\partial \alpha_j}, \\
b &\leftarrow b - \eta \frac{\partial L}{\partial b}, \\
\theta &\leftarrow \theta - \eta \frac{\partial L}{\partial \theta},
\end{align}
where \( \eta \) is the learning rate.

\section{Result}

\subsection{Data Preprocessing}
In our data preprocessing stage, we employ Part-of-Speech (POS) tagging—a fundamental task in NLP —as a benchmark for evaluating our methods. Following the methodologies outlined in prior studies \cite{paszke2019pytorch,di2022dawn}, we implement the data processing workflow using the PyTorch framework due to its flexibility and widespread adoption in the NLP community. For illustrative purposes, we select two sentences—"The dog eat the ice" and "Everybody read that book"—and manually assign POS tags to each word. Specifically, the labels for the first sentence are \texttt{["DET", "NN", "V", "DET", "NN"]}, corresponding to the POS of each word and facilitating syntactic structure analysis.

During data preparation, we first tokenize the sentences and convert the tokens into word index tensors through word indexing. This process utilizes a pre-established vocabulary where each unique token is assigned a unique index, enabling the mapping of tokens to their numerical representations required for computational processing. Subsequently, we transform the POS labels into indexed tensors via label mapping. This step allows the model to associate each POS tag with its corresponding numerical index during training, which is essential for effectively learning the underlying patterns associated with each POS tag.

\subsection{Performance Benchmmarking}

\begin{table}[!b]
    \caption{Comparison of Parameters for QK-LSTM and LSTM Networks.}
    \label{tab:quantum_lstm_vs_lstm_parameters}
    \centering
    \begin{tabular}{@{}lcc@{}}
        \toprule
        \textbf{Parameter} & \textbf{QK-LSTM} & \textbf{LSTM} \\
        \midrule
        Epochs & 100 & 100 \\
        Learning Rate & 0.1 & 0.1 \\
        Number of Tags & 3 & 3 \\
        Vocabulary Size & 5 & 5 \\
        Embedding Dimension & 8 & 8 \\
        Hidden Dimension & 6 & 6 \\
        Number of Qubits in Quantum Kernel Circuit & 4 & -- \\
        \midrule
        \textbf{Total Trainable Parameters} & \textbf{183} & 477 \\
        \bottomrule
    \end{tabular}
\end{table}

The QK-LSTM model effectively compresses the traditional LSTM architecture by leveraging quantum kernel computations, reducing the need for large embedding and hidden dimensions. As shown in Table~\ref{tab:quantum_lstm_vs_lstm_parameters}, the QK-LSTM has significantly fewer trainable parameters (183) compared to the classical LSTM (477), primarily due to the use of quantum kernel circuits that enhance feature representation without relying on extensive parameterization. This compression is achieved by encoding complex patterns and correlations within a lower-dimensional quantum Hilbert space, allowing the QK-LSTM to capture intricate dependencies with fewer parameters.

\begin{figure}[!t]
    \centering
    \includegraphics[width=0.4\textwidth]{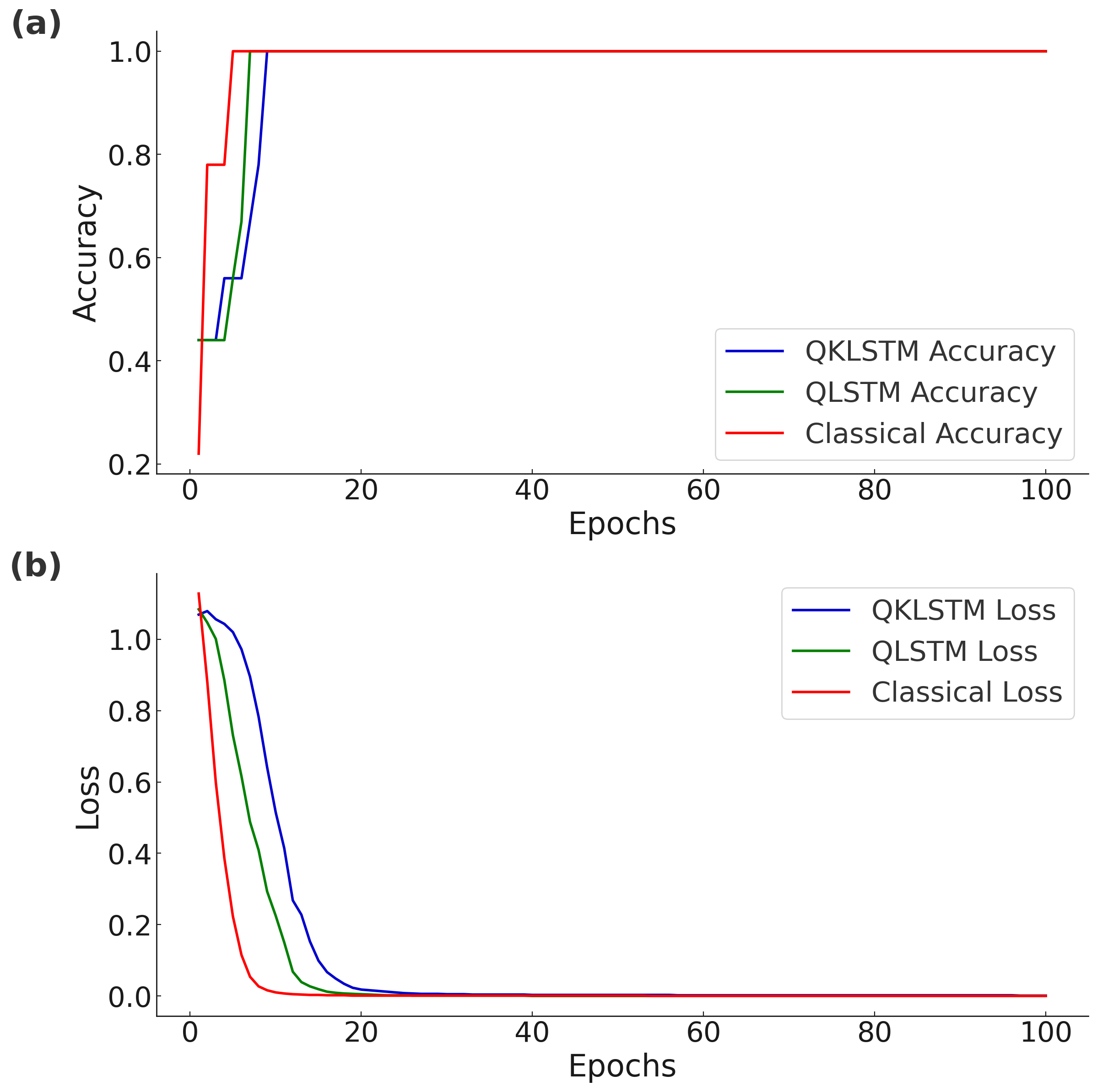}
    \caption{Training performance comparison for QLSTM, Classical, and QK-LSTM models. (a) Accuracy over epochs. (b) Loss over epochs, showing optimization trends for each model.}
    \label{result}
\end{figure}

The efficacy of this compressed model is evident in the convergence and optimization performance metrics. Fig.~\ref{result}(a) illustrates that the QK-LSTM attains accuracy levels comparable to the classical LSTM and QLSTM, with a similar rate of convergence despite the reduced parameter set. Furthermore, in  Fig.~\ref{result}(b), the QK-LSTM demonstrates robust loss minimization and stability over epochs, achieving rapid optimization akin to the more parameter-heavy classical LSTM. This efficiency suggests that quantum kernels enable the QK-LSTM to maintain a high capacity for representation while minimizing resource demands, leading to a compact and computationally efficient architecture. Such model compression is advantageous for real-world applications where memory and processing constraints are critical, highlighting the QK-LSTM's potential for deployment in edge computing environments or devices with limited computational power.

\section{Discussion}
The QK-LSTM model demonstrates significant strides in the application of quantum-enhanced machine learning by effectively incorporating quantum kernel functions within a classical LSTM architecture. This integration not only leverages quantum feature spaces to capture intricate data dependencies with fewer parameters but also achieves model compression without sacrificing accuracy. The QK-LSTM’s performance underscores the potential of quantum kernels to enhance computational efficiency, making it particularly suitable for deployment in resource-constrained environments, such as edge devices. Benchmark comparisons with traditional LSTM networks illustrate that QK-LSTM maintains competitive accuracy and convergence rates while minimizing resource demands, highlighting its practicality in real-world applications where memory and processing power are limited. The findings suggest that quantum kernel methods hold considerable promise in advancing QML, offering a viable pathway to develop efficient and scalable models that bridge current hardware constraints.



\clearpage

\bibliographystyle{ieeetr}
\bibliography{reference}

\vspace{12pt}

\end{document}